\documentclass[fleqn,twoside,epsf]{article}
\usepackage{espcrc2}

\usepackage{graphicx}

\newcommand{\AmS}{{\protect\the\textfont2
  A\kern-.1667em\lower.5ex\hbox{M}\kern-.125emS}}

\newcommand{\be }{\begin{equation}}
\newcommand{\ee }{\end{equation}}
\newcommand{\beq}{\begin{eqnarray}}
\newcommand{\eeq}{\end{eqnarray}}
\title{The pentaquark potential, mass and  density-density correlator}

\author{C.Alexandrou\address[Cyprus]{Department of Physics,
University of Cyprus, CY-1678 Nicosia, Cyprus}, G.Koutsou\addressmark[Cyprus]
 and A.Tsapalis\thanks{Supported by the Levendis Foundation.}\addressmark[Cyprus]
}

\begin{document}

\begin{abstract}
We evaluate the static $qqqq\bar{q}$
 potential in the quenched theory
at $\beta=5.8$ and $\beta=6.0$ on a lattice of size $16^3\times 32$. 
The mass and density-density correlator for the $\Theta^+$ is investigated in the quenched theory
at $\beta=6.0$ on lattices of size $16^3\times 32$, $24^3\times 32$ and $32^3 \times 64$.
\end{abstract}

\maketitle

\vspace*{-1.5cm}

\section{Introduction}

A large amount of effort is being devoted  to experimental      
searches
for the identification  of the $\Theta^+$, an exotic baryon state with an unusually  narrow width. 
The possible existence of such a state has
 raised interesting questions about its structure.
A number of phenomenological models have been put forward to explain its stability 
such as special flux tube formation~\cite{nussinov} and diquark formation~\cite{Jaffe}.
The focus of lattice studies has been the calculation of the $\Theta^+$ mass 
and the identification of its
parity~\cite{lattice,Liu}.
In this work we look at  the density-density correlator
which yields information on the quark distribution inside a hadron.
In addition we evaluate  
the static pentaquark potential by 
constructing the pentaquark Wilson loop.
We compare the  static pentaquark  potential to the potential extracted in the strong coupling approximation
as well as to the sum of the baryonic and  mesonic potentials. 

\vspace*{-0.3cm}

\section{Static potential}
\begin{figure}[h]
\vspace*{-1cm}
\centerline{\mbox{\includegraphics[height=4cm,width=5.cm]{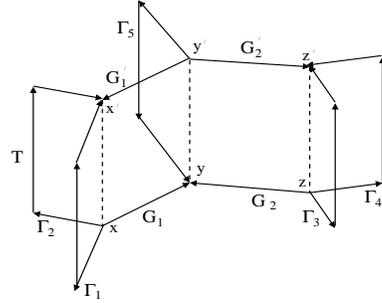}}}
\vspace*{-1cm}
\caption{The Wilson loop , $W(T)$.}
\label{fig:pentaq_loop}
\end{figure}

\begin{figure}[h]
\vspace*{-1cm}
\begin{minipage}[t]{3.5cm}
\centerline{\mbox{\includegraphics[height=3.5cm,width=3.5cm]{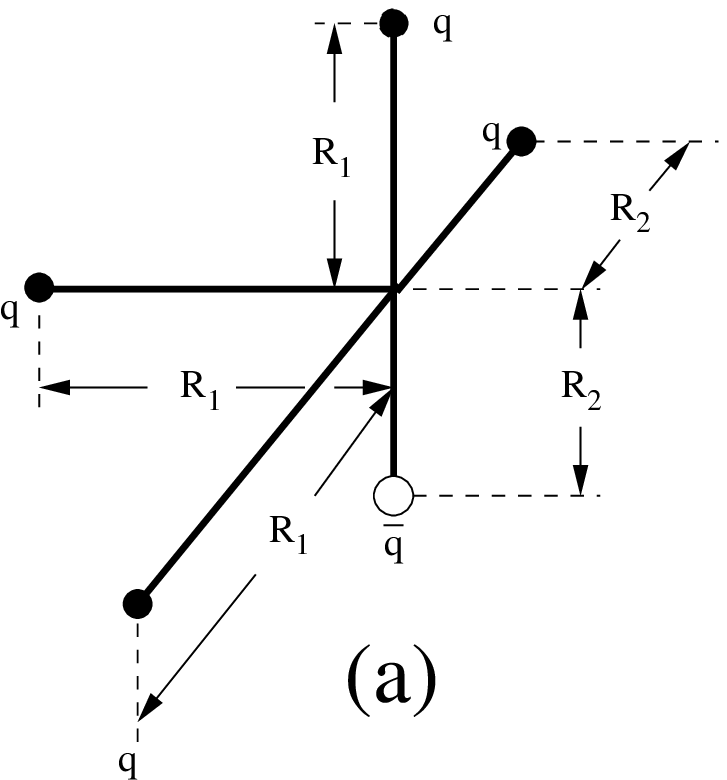}}}
\end{minipage}
\begin{minipage}[t]{3.5cm}
\centerline{\mbox{\includegraphics[height=3.5cm,width=3.5cm]{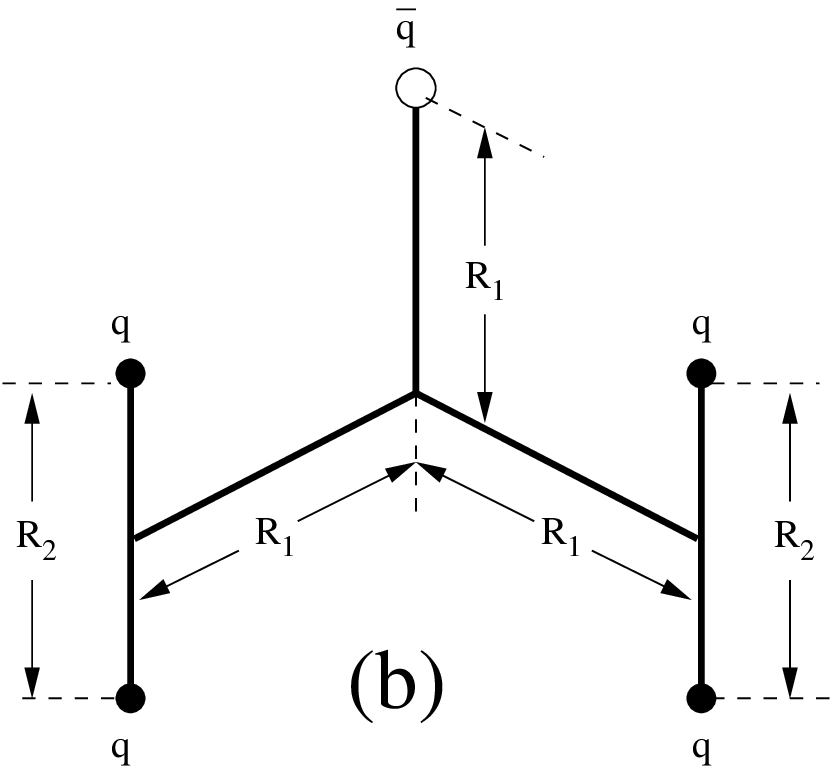}}}
\end{minipage}
\vspace*{-1.1cm}
\caption{(a) Geometry I, (b) Geometry II}
\vspace*{-0.9cm}
\label{fig:pentaq_geom}
\end{figure}

The SU(3)  Wilson loop for the pentaquark is shown in Fig.~\ref{fig:pentaq_loop}: it is constructed 
by creating a gauge invariant
$qqqq\bar{q}$ quark state at time $t=0$ which is annihilated at a later time $T$~\cite{paper1}.
We consider  two  geometries as shown 
in Fig.~\ref{fig:pentaq_geom}. Geometry I is based on
 the KN structure whereas geometry II on the diquark structure probed as 
a function of the distance $R_2$.
The potential is extracted by 
fitting the ratio  
$ -ln(W(t+1)/W(t))$
in the plateau region.
We use multi-hit on the temporal links and 30 levels of APE smearing on the
spatial links with smearing weight $\alpha=1/2$.

\begin{figure}[h]
\vspace*{-0.8cm}
\centerline{\mbox{\includegraphics[height=4.cm,width=7cm]{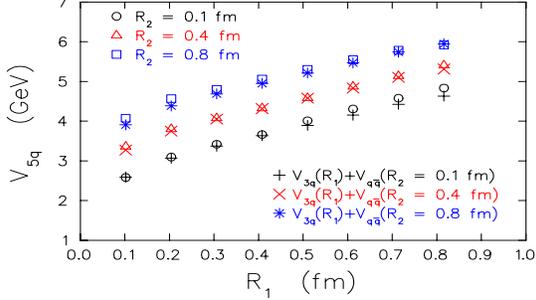}}}
\vspace*{-0.8cm}
\caption{The pentaquark static potential for geometry I for $R_2$=0.1, 0.4 and
0.8~fm.
The crosses, the x-symbols and the stars are lattice data for the sum of
the baryonic and mesonic potential at  $R_2 = 0.1, 0.4, 0.8$~fm respectively.}
\label{fig:geometryI}
\vspace*{-0.8cm}
\end{figure}

\begin{figure}[h]
\vspace*{-0.8cm}
\centerline{\mbox{\includegraphics[height=4.5cm,width=8cm]{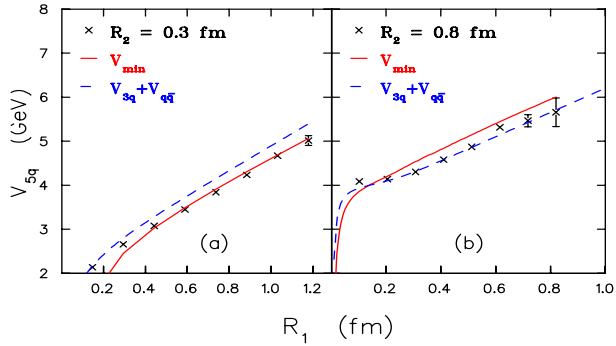}}}
\vspace*{-0.8cm}
\caption{The pentaquark static potential for geometry II:
  (a) for $\beta=5.8$  and $R_2 = 0.3$ fm, 
(b) for $\beta=6.0$  and  $R_2 = 0.8$ fm.
The solid line is $V_{\rm min}$ and the dashed line the sum of the  baryonic 
and  mesonic potentials.}
\label{fig:geometryII}
\vspace*{-1cm}
\end{figure}

All the computations are carried out on a lattice of size $16^3\times 32$
at $\beta=5.8$ and $6.0$ using 200 configurations
available at the NERSC archive. Comparison  of data at these two $\beta$-values
shows that the potential has good scaling properties~\cite{paper1}. 
The results for the pentaquark potential using 
geometry I are shown in Fig.~\ref{fig:geometryI}. 
On the same figure we also show lattice data for the sum of the corresponding
 baryonic potential, $V_{3q}(R_1)$,
 and  the mesonic $V_{q\bar{q}}(R_2)$. 
As can be seen the pentaquark potential is the same as 
the potential of the KN system.
In Fig.~\ref{fig:geometryII} we show the pentaquark potential using
geometry II for two very different values
of the distance $R_2$:
1) For $R_2=0.3$~fm, which is  the smallest possible separation at $\beta=5.8$,
the potential for $R_1>R_2$ is well described by 
\be 
V_{\rm min}^{5q} (L_{\rm min})=
\frac{5}{2}  V_0-n_q\sum_{i> j}\frac{\alpha}
{|{\bf r}_i-{\bf r}_j|}+\sigma L_{\rm min}
\label{Vmin}
\ee
where $L_{\rm min}$ is the minimal length joining the quarks.
 $V_0, \alpha$ and $\sigma$ are extracted from fitting
the $q\bar{q}$ potential to the form 
$V_{q\bar{q}}(r)=V_0 -\frac{\alpha}{r} + \sigma r$
and the factor $n_q$ in front of the Coulomb term  is one between $q$ and 
$\bar{q}$
and 1/2  between (anti-)quarks  as obtained 
from one-gluon exchange.
 This genuine pentaquark state has   
static energy which is lower than the sum of the baryonic 
and  mesonic potentials.
2) For
$R_2=0.8$~fm, which is the largest possible separation  at $\beta=6.0$ 
and for which 
$R_1\le R_2$,
 the results are 
well described by the sum of the baryonic and  mesonic potentials 
and only for larger distances 
they tend to 
 approach $V_{\rm min}$.
  For this comparison 
the baryonic potential
is parametrized using the $Y-$ Ansatz:
$\frac{3}{2}  V_0-\frac{1}{2}\sum_{i> j}\frac{\alpha}
{|{\bf r}_i-{\bf r}_j|}+\sigma L_{\rm min}$.

\section{Mass and density - density correlator}

\begin{figure}[h]
\vspace*{-1cm}
\centerline{\mbox{\includegraphics[height=2.8cm,width=5cm]{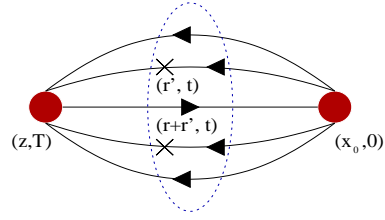}}}
\vspace*{-1cm}
\caption{Density-density correlator.}
\label{fig:theta}
\vspace*{-0.8cm}
\end{figure}

The purpose  is to study the distribution of 
quarks inside the $\Theta^+$ given that we can identify it on our lattices.
The quark distribution can be studied via
the  density-density correlator  shown in Fig.~\ref{fig:theta}
and given by~\cite{APT}
\be
C_{uf}({\bf r},t) = \int d^3r' \> <\Theta | j_0^u({\bf r}+{\bf r}^{\prime},t)
 j_0^f({\bf r},t)|\Theta>
\label{dens-dens}
\ee
where $j_0^f({\bf r},t)=:\hspace*{-0.1cm}\bar{f}({\bf r},t)\gamma_0 f({\bf r},t)\hspace*{-0.1cm}:$ for quark of
flavour $f$.
In the non-relativistic limit  it reduces to the wave function squared.
Therefore the correlator probes the $\Theta^+$ wave function in a gauge-invariant way
unlike Bethe - Salpeter amplitudes.

We use Wilson fermions and lattices of size $16^3\times 32$, $24^3 \times32$ 
and $32^3\times64$ 
at $\beta=6.0$.
We impose Dirichlet (anti-periodic) boundary conditions 
in the temporal direction for the
lattices of temporal size $32$ (64).
We fix  ${\kappa}_s=0.155$ for the strange quark propagator. 
This choice gives  
${m_k}/{m_N}=0.5$ and ${m_\phi}/{m_N}=1.04$ close to the experimental ratios.
The light quark propagators are computed at 
 $\kappa_l=$0.153, 0.154, 0.155, 0.1554, 0.1558 and 0.1562
 $({m_\pi}/{m_\rho} =  0.83,  0.78, 0.70, 0.64, 0.58, 0.50)$ 
 with the two largest done only on 
 the $32^3\times64$ lattice. 
 For the $\Theta^+$ interpolating  field we use the diquark-diquark combination:
\be
\Theta^+ =\epsilon^{abc}\epsilon^{aef}\epsilon^{bgh} \> C\bar{s}_c^T 
\left(u_e^T C d_f\right) \left(u_g^T C \gamma_5 d_h\right)
\label{theta}
\ee
 
\begin{figure}[h]
\vspace*{-0.5cm}
\centerline{\mbox{\includegraphics[height=4.cm,width=8cm]{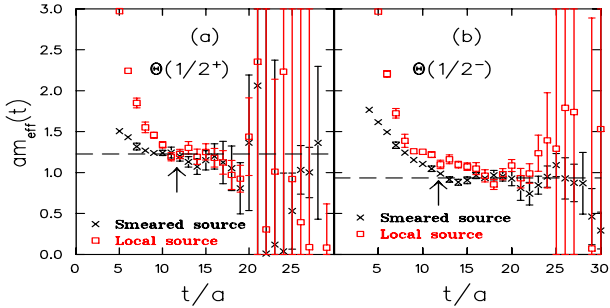}}}
\vspace*{-0.8cm}
\caption{$\Theta^+$ effective mass, $m_{\rm eff}(t)$, on the  $32^3\times 64$
lattice at $\kappa_l=0.1554$:
(a) for positive  and (b) for negative parity. 
Arrows show the time slice where the density operators are inserted.}
\label{fig:theta_0.1554}
\vspace*{-0.9cm}
\end{figure}

Our effective masses  for a Wuppertal smeared  and local source
are shown in Fig.~\ref{fig:theta_0.1554} 
for $\kappa_l=0.1554$. 
For the $\Theta(1/2^+)$ the plateaus coincide for $t/a>10$ whereas for
$\Theta(1/2^-)$ only when $t/a>15$. Beyond this time separation only one
plateau can be identified.
 In order to establish whether in this time range we have a bound 
state we compare the volume dependence of the weights in the two-point 
function~\cite{Liu} using local source and  sink. We fit 
in the range (11-15) for the $\Theta(1/2^+)$ and in the range (16-22) for
the $\Theta(1/2^-)$ where smeared and local sources give the same plateaus.
Denoting by $w_L$ the weight of the state on a lattice of spatial
size $L$ we find at $\kappa_l=0.153$ 
for the $\Theta(1/2^+)$ $w_{16}/w_{24}=0.42(5)$, $w_{16}/w_{32}=0.71(9)$
and $w_{24}/w_{32}=1.64(18)$ 
and for the $\Theta(1/2^-)$ $w_{16}/w_{24}=1.17(20)$, $w_{16}/w_{32}=0.56(10)$
and $w_{24}/w_{32}=0.48(5)$ where the errors are only statistical i.e.
evaluated within a
one parameter fit  of the correlators giving as
input the mass
extracted
 from first fitting $m_{\rm eff}(t)$ keeping the fitting ranges the same.
 Similar values are
obtained at $\kappa_l=0.1554$. If the state is a scattering state the
ratios should scale like the spatial volume
i.e $w_{16}/w_{24}=3.4$, $w_{16}/w_{32}=8$
and $w_{24}/w_{32}=2.4$. The values for the ratios that we find are
 closer to one
supporting single particle states. Extracting the mass from 
 lattice data with
a smeared source  at $\kappa_l=0.1554, 0.1558$ and 0.1562 on the 
$32^3\times 64$ lattice and extrapolating to the
chiral limit we find for the mass ratios $m_{\Theta(1/2^+)}/m_N=2.75(12)$ and 
$m_{\Theta(1/2^-)}/m_N=1.57(3)$ where $m_N$ is the nucleon mass.

The density operators are inserted at $t/a=12$, which is within the plateau
range for both parity states when a smeared source is used.
We show the results for $\kappa_l=0.1554$ obtained with 
100 configurations on the
$32^3\times64$ lattice.
Comparison of $C_{ud}(r)$ and $C_{us}(r)$ for $\Theta(1/2^+)$ and
$\Theta(1/2^-)$ shows that the u-s quark 
distribution is broader than the u-d distribution in particular  for 
the $\Theta(1/2^-)$.

\begin{figure}[h]
\vspace*{-0.5cm}
\centerline{\mbox{\includegraphics[height=4.cm,width=8cm]{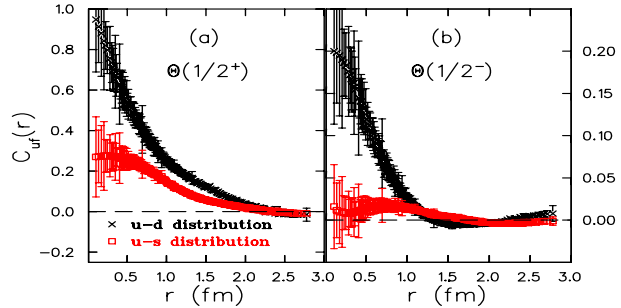}}}
\vspace*{-0.8cm}
\caption{$C_{ud}$ (black) and $C_{us}$ (red) for  $\Theta^+$  scaled by an overall constant
at $\kappa_l=0.1554$: (a) for positive and (b) for  negative parity.}
\vspace*{-1cm}
\label{fig:wf_theta_ud_us}
\end{figure} 

\vspace*{-0.3cm}

\section{Conclusions}

\vspace*{-0.3cm}

The pentaquark potential for geometries that 
favour diquark formation  is well described by $V_{\rm min}$
as given in Eq.~(\ref{Vmin}).
Otherwise 
 the potential is closer to the sum
 of the baryonic and mesonic potentials. 
The ratio of 
weights of the $\Theta^+$ correlators on spatial volumes $16^3$, $24^3$
and $32^3$ are closer to one  indicating a single particle state. 
Extrapolating to the chiral limit we find that the mass of the $\Theta(1/2^+)$
is 2.56(11)~GeV and the mass of  $\Theta(1/2^-)$ is 
1.46(3)~GeV where we used  the nucleon mass to convert to physical units. 
However one has to keep in mind that the lower 
KN scattering states can not be clearly  identified in the current analysis. 
The density-density correlators show 
that in both  parity states of the $\Theta^+$
the u-s quark distribution is broader than the u-d distribution. This 
difference is
particularly striking 
 in the case of the  
$\Theta(1/2^-)$.

\end{document}